# False metals, real insulators, and degenerate gapped metals

Oleksandr I. Malyi and Alex Zunger

Renewable and Sustainable Energy Institute, University of Colorado, Boulder, Colorado 80309, USA

E-mail: Alex.Zunger@colorado.edu

This paper deals with a significant family of compounds predicted by simplistic electronic structure theory to be metals but are, in fact, insulators. This false metallic state has been traditionally attributed in the literature to reflect the absence of proper treatment of electron-electron correlation ("Mott insulators") whereas, in fact, even mean-field like density functional theory describes the insulating phase correctly if the restrictions posed on the simplistic theory are avoided. Such unwarranted restrictions included different forms of disallowing symmetry breaking described in this article. As science and technology of conductors have transitioned from studying simple *elemental* metals such as Al or Cu to *compound* conductors such as binary or ternary oxides and pnictides, a special class of degenerate but gapped metals has been noticed. Their presumed electronic configurations show the Fermi level inside the conduction band or valence band, yet there is an "internal band gap" between the principal band edges. The significance of this electronic configuration is that it might be unstable towards the formation of states inside the internal band gap when the formation of such states costs less energy than the energy gained by transferring carriers from the conduction band to these lower energy acceptor states, changing the original (false) metal to an insulator. The analogous process also exists for degenerate but gapped metals with Fermi level inside the valence band, where the energy gain is defined by transfer of electrons from the donor level to the unoccupied part of the valence band. We focus here on the fact that numerous electronic structure methodologies have overlooked some physical factors that could stabilize the insulating alternative, predicting instead false metals that do not really exist (note, this is in general not a physical phase transition but a correction of a previous error in theory that led to a false prediction of a metal). Such errors include: (i) ignoring *spin symmetry breaking*, such as disallowing magnetic spin ordering in $CuBi_2O_4$ or disallowing the formation of polymorphous spin networks in paramagnetic $LaTiO_3$ and $YTiO_3$; (ii) ignoring *structural symmetry breaking,* e.g., not enabling energy-lowering bond disproportionation (Li-doped $TiO_2$, $SrBiO_3$, and rare-earth nickelates) or not exploring pseudo-Jahn-Teller-like distortions in $LaMnO_3$ or disallowing spontaneous formation of ordered vacancy compounds in $Ba_4As_3$ and $Ag_3Al_{22}O_{34}$; (iii) ignoring spin-orbit coupling (SOC) forcing false metallic states in $CaIrO_3$ and $Sr_2IrO_4$. The distinction between false metals vs real insulators is important because, (a) predicting theoretically that a given compound is metal even though it is found to be an insulator often creates the temptation to invoke a high order novel physical effects (such as correlation in d electron Mott insulators) to explain what was in effect caused by a more mundane artifact in a lower-level mean-field band theory, (b) recent prediction of exotic physical effects such as topological semimetals were unfortunately based on the above compounds that were misconstrued by theory to be metal, but are now recognized to be stable insulators not hosting exotic effects, and (c) practical technological application based on stable degenerate but gapped metals such as transparent conductors or electrides for catalysis must rely on the systematically correct and reliable theoretical classification of metals vs insulators.

# I. Introduction

One of the most fundamental descriptors of solids is their designation as metals or insulators. Indeed, this distinction frames much of the discussion of their electronic, transport, superconducting, or topological characteristics.[1-4] In standard theoretical descriptions, this distinction is represented via the construct of energy vs wavevector (E vs k) band structure, where metals (respectively, insulators) have their Fermi level ($E_F$) inside the continuous part of the band structure (respectively, band gap) energy region. Figure 1a shows schematically the density of states of an n-type metal where the Fermi level resides inside the conduction band (CB). This is often the case when the composition weighted formal oxidation state (FOS) is positive, as is the case for $Sr^{2+}V^{5+}O_3^{2-}$.[5] A p-type metal would correspond to the case where the Fermi level resides in the valence band (VB), *as is often the case when the composition weighted FOS is negative, e.g.,* $Tl^{1+}Cu_2^{1+}Se_2^{2-}$.[6] Unlike the band structure of *elemental metals* such as Al or Cu, where the next occupied band below the metallic band is generally an isolated, electronically inactive *core electron band*, recent interest in *compound metals* (such as the familiar groups $ABX_2$, $A_2BX_4$, and $ABX_3$ ternary chalcogenides and pnictides some being metallic), has focused attention on the special electronic configuration illustrated in Fig. 1a, namely, where a principal occupied valence band is located just a couple of eV below the metallic conduction band. We will refer to a metallic system having its Fermi energy inside the conduction or valence band, yet and a large "internal band gap" ($E_g^{int}$ in Fig. 1a) between the principal valence band maximum and conduction band minimum as a *"degenerate but gapped metal"*.

These prototypical electronic configurations are well known in *dilute* doped inorganic and organic insulators.[7-11] Interestingly, there are also many *pristine (undoped) stoichiometric solids*, where the electronic structure corresponds to the configuration of degenerate gapped metal portrayed in Fig. 1a. The most famous examples of such compounds are $SrVO_3$[5], $CaVO_3$[5], $BaNbO_3$[12], and $Ca_6Al_7O_{16}$[13], which exhibit metallic behavior based on the analysis of temperature behavior of resistivity[5,13,14] and angle-resolved photoemission spectroscopy.[15-17] The successful design of the novel degenerate gapped metals lies at the heart of the development of transparent conductors[18], electrides[19], and prediction of Dirac semimetals.[20-22] For instance, it has been recently proposed that intrinsic degenerate gapped metals can be used as transparent conductors.[23,24] Similarly, many electride[13] compounds correspond to bulk solids that have degenerate gapped electronic structures where the electrons form a two-dimensional (2D) or 1D or 0D cloud resembling an intrinsic electron gas.

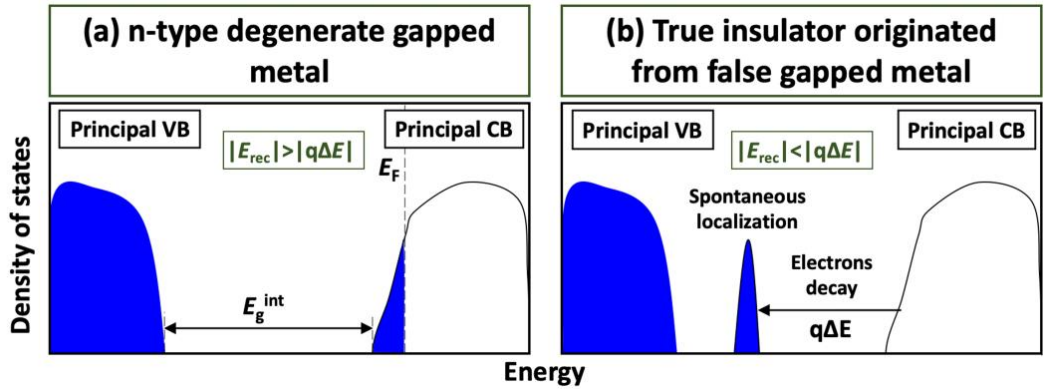

**Figure 1.** Schematic illustration of (a) degenerate gapped metal having electrons in the principal conduction band and, in addition, an "internal" band gap $E_g^{int}$ between the principal conduction and valence bands; (b) Schematic illustration of real insulator originating from a degenerate gapped metal. Blue: occupied states, white: unoccupied states. Here, the system invests energy $E_{rec}$ to create in the gap a new state via reconstruction, magnetization, or defect formation. In turn, this state accepts q electrons from the conduction band, contributing to energy lowering $q\Delta E$, where $\Delta E$ is the difference in band energies for electrons in the conduction band and localized occupied state. This creates a gap between the newly formed state and principal conduction band. The figure is reproduced with permission from Malyi and Zunger Phys. Rev. B, 101, 235202 (2020), Copyright (2020) by the American Physical Society.

***Spontaneous instabilities:*** Free carriers introduced via doping to insulators are the source of "doping bottlenecks",[25-27] whereby doping leads to the self-regulating formation of *charged* structural defects that compensate the intentional doping. But such free carrier instabilities can also occur in pristine compounds having analogous electronic configurations. Figure 1b illustrates the fact that such electronic structure configuration might be unstable towards the formation of an "acceptor state" inside the principal band gap if its formation costs a "reconstruction (rec) energy" $|E_{rec}|$ that is smaller than the energy gained by transferring q electrons from the conduction band to these lower energy acceptor states $|q\Delta E|$, where $\Delta E$ is the band energies difference for electrons in conduction/valence band and localized states. In general, spontaneously formed acceptor states may be *structural defects* (e.g., cation vacancies) as in $BaNbO_3$, $Ca_6Al_7O_{16}$, and $Ag_3Al_{22}O_{34}$ and leading to (i) the observed off stoichiometry even at low-temperatures[23,28-31], (ii) reduced metallicity and, at the extreme, even to (iii) the emptying of the conduction band and thus a metal to insulator transition. But such "acceptor states" may also be *electronic defects*, rather than structural defects such as polaron that can sweep conduction electrons into its "acceptor" state.[32-34] On the other hand, when $|E_{rec}|>|q\Delta E|$ in Fig. 1a, the metallic configuration can be stable against reconstructions, and the gapped metal configuration is real. The analogous process illustrated in Fig. 1b for n-type solids exists for degenerate but gapped metals that have their Fermi level inside the valence band.

***The false metal syndrome:*** It is our observation that current literature sometimes reflects confusion between false metals, real insulators, and degenerate gapped metals. We will discuss in the current paper when predictions of metallic configuration, such as that shown in Fig. 1a are valid, and when they are false, meaning that the actual electronic structure is that illustrated in Fig. 1b. The distinction between false metals vs real metals vs real insulators is important because of a number of reasons. Indeed, predicting theoretically that a given compound is metal even though it is found to be an insulator often creates the temptation to invoke a high order novel physical effects (such as correlation in d-electron compounds[35-37]), to explain the misassignment, rather than searching for a more mundane artifact in a lower-level mean-field band theory. Examples include some Mott insulators such as 3d oxides, where historically naïve band theory[35] using nonmagnetic spin configuration and a minimal unit cell forced, for an odd number of electrons per cell, a false metallic state, in contrast with experiment. This sharp contradiction with the experiment prompted the long-lasting tradition of a Hubbard Hamiltonian description of false metals would be Mott insulators that identify many-body correlation effects as the gapping mechanism. However, a number of examples came to light illustrating that ordinary density functional theory (DFT) free from the restrictions to the nonmagnetic description, or its commitment to minimal unit cell sizes already correctly gives an insulating ground state.[38-42] This raises the question of what type of physics is responsible for the removal of false metal designation in favor of real insulators. This will be discussed in the present paper.

While recently developed open access databases[20,43-45] of electronic band structures constitute an important contribution to materials science, false metals abound such databases (see a few examples in Fig. 2) and literature citing them discussed below. Indeed, it was recently argued that the prediction of novel Dirac metal $(BiO_2)$[46] is unlikely to be valid, as a more proper description[31] of the same compound shows that it spontaneously reconstructs to significantly lower energy creating a trivial insulator structure. Hence, herein, we will show that the leading cause of the *"false metal syndrome"* is the incomplete application of electronic structure theory, omitting a degree of freedom in the calculation that, if enabled, will convert the false metal into a real insulator, as illustrated schematically in Fig. 1b.

| Compound | Materials project | OQMD | Topological Materials Database | AFLOW | Experiment |
|---|---|---|---|---|---|
| NiO | *Insulator* | *Insulator* | Metal | Insulator | Insulator |
| MnO | *Insulator* | *Insulator* | Metal | Insulator | Insulator |
| CoO | Metal | Metal | Metal | Metal | Insulator |
| FeO | Metal | Metal | Metal | Metal | Insulator |
| $Ce_2O_3$ | Metal | *Insulator* | Metal | Insulator | Insulator |
| $Ti_6O_{11}$ | Metal | Metal | Metal | Insulator | Insulator |
| $CaIrO_3$ | Metal | Metal | Metal | Metal | Insulator |
| $Sr_2IrO_4$ | Metal | Metal | Metal | Metal | Insulator |
| $CuBi_2O_4$ | Metal | Metal | Metal | Insulator | Insulator |
| $LaTiO_3$ | Metal | Metal | Metal | Metal | Insulator |
| $YTiO_3$ | Metal | Metal | Metal | Insulator | Insulator |
| $LaVO_3$ | *Insulator* | *Insulator* | Metal | Insulator | Insulator |
| $CaVO_3$ | *Insulator* | *Insulator* | Metal | Insulator | Metal |
| $SrVO_3$ | *Insulator* | Metal | Metal | Metal | Metal |
| $YNiO_3$ | Metal | Metal | Metal | - | Insulator |
| $SrBiO_3$ | Insulator | Insulator | Insulator | - | Insulator |
| $SmNiO_3$ | Metal | Metal | Metal | - | Insulator |
| $Ba_4As_3$ | Metal | - | - | - | Insulator |

**Figure 2.** Examples of computational data for false metals available in Materials Project[43], Open Access Materials Database (OQMD)[45], Automatic - FLOW for Materials Discovery (AFLOW)[44], and Topological Materials Database[20] with comparison to experimental data. The summary demonstrates that theoretically predicted electronic structures are often in disagreement with corresponding experimental data. The reasons for the false metal predictions are not absence of electron-electron correlation but generally disallowing some modes of symmetry breaking in DFT. The cases where the insulating nature of compounds has been predicted in the respective literature from calculations of density of states (not from band structure calculations) are shown in italic. Since results in the databases are changing with time, we note that the data quoted here was retrieved on May 28, 2020.

Figure 3 summarizes the modalities that can lead to the prediction of false metals and will be discussed in the current article. The first category involves oversimplified *computational assumptions such as* restriction of the unit cell representation to cells that cannot geometrically accommodate low symmetry structures or use of exchange- correlation (XC) functionals that produce rather delocalized orbitals unable to take advantage of energy lowering broken symmetries. The second category involves *oversimplified physical assumptions* that restrict the formation of structural, magnetic, or defect breaking modes that could otherwise transform configuration Fig. 1a to Fig. 1b. In this case, the initial configuration in Fig. 1a exists only under hypothetical, theoretical approximations that fail to take advantage of this energy lowering reconstruction.

The paper is organized as follows. Sec. II provides basic information on the different computational approximations for describing electronic structures of compounds, which can result in the prediction of false metals. Sec. III gives details on symmetry breaking motifs that should be accounted for during the search of potential degenerate gapped metals providing detailed examples for each case. In Sec IV, we illustrate the cases when false metals can create in-gap polaron-like states and their importance. In Sec. V, we discuss the progress on the theoretical and experimental prediction of real degenerate gapped metals. Finally, we provide a short outlook and perspective of the field in Sec. VI.

| Cause of False Metal | | Cause of gapping false metal | Example |
|---|---|---|---|
| Computational | Insufficient constraints on XC functional | XC distinguishing occupied form unoccupied states | All systems |
| | | XC which reduces self interaction error | |
| | Restricting unit cell disallowing symmetry breaking | Flexible representation for the unit cell | |
| Symmetry breaking motifs | Ignoring local magnetic motifs | Magnetic order | $CuBi_2O_4$, NiO, all magnets below |
| | | Different local spin environments | PM $LaTiO_3$, PM $YTiO_3$, PM NiO, PM $YNiO_3$ |
| | Ignoring local structural/orbital motifs | Octahedral tilting | $SrBiO_3$, $BaBiO_3^*$, $Li(TiO_2)_{16}$, $YNiO_3^*$, $SmNiO_3^*$ |
| | | Atomic displacement | |
| | | Disproportionation | |
| | | $Q_2^+$ distortion | Cubic $LaMnO_3^*$ |
| | Ignoring defect induced symmetry breaking | Defect formation | $Ba_4As_{3-x}$, $Ag_3Al_{22}O_{34}$ |
| | Ignoring spin-orbit coupling | Allowing spin-orbit coupling | $CaIrO_3$, $Sr_2IrO_4$ |

**Figure 3.** Different causes of false metals and mechanisms of the corresponding band gap opening with examples. Compounds that are observed to be metals at high temperature are marked by a star.

## II. The general framework of theory that permits coupling between atomic structure, spin configuration, and electronic properties

Different styles of electronic structure theories have different vulnerabilities towards predicting false states of conductivity. *Fixed-Hamiltonian band structure methodologies* (see textbooks[1-4]) lack a feedback mechanism that allows the electronic structure and the crystal or spin structure to affect each other. This is the case for standard tight binding[47], k·p[48,49], or empirical pseudopotential method[50,51], that selected fixed atomic compositions, crystal structure with specific internal symmetry, unit cell size, and a fixed spin configuration that are not allowed to vary as the electronic structure does. These fixed features end up deciding uniquely the outcome electronic characteristics – whether true or false.

Modern electronic structure modalities based on DFT, on the other hand, allow the electronic structure to respond to the different occupation numbers, atomic positions, spin configurations, unit cell size, and symmetries, variables that are being explored during the calculation in search of minimum energy self-consistent solution. This provides a feedback loop between *the dynamic variables of the calculation* **I** – {occupation numbers, atomic positions, spin configuration, cell symmetries} – all able to change during the solution process, and the ensuing electronic structure **II** (stable atomic and magnetic structure, energies of states, and their localization). The *agents* mediating the effects of the variables of the calculation **I** on the ensuing electronic structure **II** are the well-known heuristic constructs of traditional solid-state chemistry such as bonding, charge transfer and hybridization. An important manifestation of this feedback response, for example, is that electron addition (to empty states) and electron removal (from occupied states) can change the underlying difference in their energies, i.e., the band gap, transforming, for example, a false metal to a real insulator. We will explain in this article that *this crucial nonlinear feedback between I and II is a property already present in properly formulated mean-field DFT (but not of unresponsive fixed-Hamiltonian band structure methodologies noted above)*, rather than the exclusive property *of many-body dynamic correlation effects, an opinion as very often echoed in the literature.* Thus, compounds such as NiO or Cuprate superconductors that were depicted as false metals in the literature using naïve, fixed Hamiltonian band structure models, became correctly insulators using properly executed symmetry-broken DFT. The present article will illustrate numerous mechanisms (Fig. 3) where the above noted feedback loop underlying mean-field theories depicts correct symmetry broken insulation without recourse to strongly correlated symmetry preserving treatment.

Figure 4 illustrates the basic makeup of such electronic structure methods. For a given compound defined by its **A**tomic identities, **C**omposition, and **S**tructure type (ACS for short[52]), there are two types of inputs: one that is fixed (the inner circle) and one that changes iteratively during the calculation (outer circle). *The fixed input* contains the definition of the compound explored, i.e., ACS. This is encoded in the atomic numbers, pseudopotential, and the crystalline phase of interest. These quantities appear in the crystal electron-ion potential $V_{ext}(r)$, which is the changing part of the input. In addition, the fixed part also contains the type of screening $V_{scr}(r)$ allowed for the external potential. In the context of DFT, this refers to the exchange-correlation potential.[53] *The changing part* consists of parameters that are affected by the evolving electronic structure and need to be optimized as the calculation goes forwards.

This includes energy-minimizing structural parameters consistent with the physical phase ACS being explored. Such parameters include the self-consistent charge and spin densities, atomic position parameters obtain via minimization of atomic forces, the size of cell (primitive or supercell), and type of spin configuration. These quantities can change during the calculation in order to find the lowest total energy. Different levels of computational sophistication do this, i.e., either automatic variations seeking lower total energy[54-57] or "manually", exploring discrete geometries.

This framework allows coupling (i) atomic structure and spin configuration to the (ii) charge density and potential, and hence can change the ensuing band structure dramatically as the (i)-(ii) feedback unfolds iteratively during self-consistency and ionic relaxation. Herein, the calculations are performed using a plane-wave pseudopotential density functional method as implemented by the Vienna Ab Initio Simulation Package (VASP)[58-60] with the results visualized using Vesta[61]. As we will see in this article, this strong nonlinear interrelation between structure vs band structure brings new possibilities of classification and, indeed, misclassification of compounds as metals or as insulators, depending on how this coupling is described.

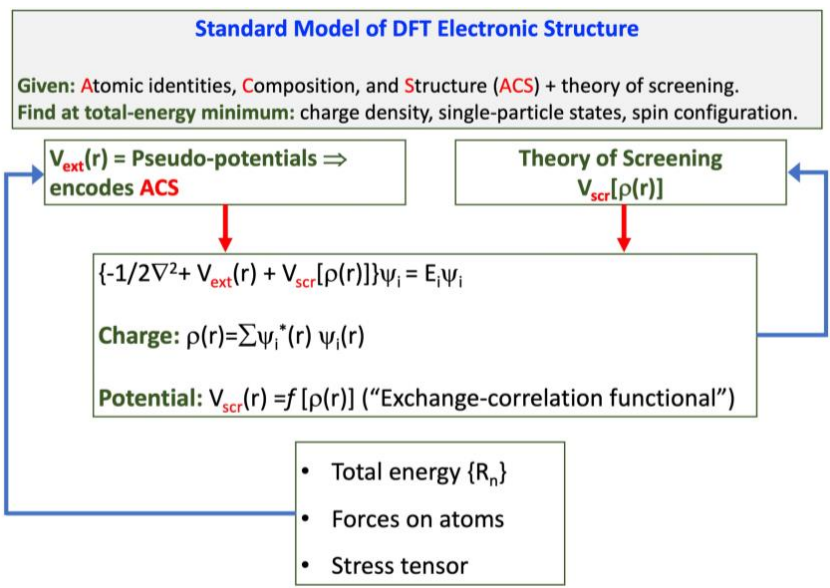

**Figure 4.** Schematic illustration of electronic structure calculations using two types of input to establish the single-particle Schrodinger equation. The inner circle is the self-consistency of the charge density for *given* $V_{ext}(r)$, and *given* spin configuration, whereas the outer circle involves *changing the geometry* [i.e., $V_{ext}(r)$] and *the spin configuration* in search of lower total energy.

Given the computational framework described in Fig. 4 that permits validation, we can now return to Fig. 3 and examine the type of *approximations* beyond what Fig. 4 afford, checking which of these additional approximations limits the full extent of coupling between electronic structure and atomic structure/spin configuration, thus leading to false metal designations.

### A. Computational need: Proper choice of the exchange-correlation functional

From the computational DFT perspective, the description of the electronic structures of compounds requires the utilization of the XC functional depicted in Fig. 4 as $V_{scr}$. There are, by now, a large number of candidate XC functionals[62], fitting different aspects. But recently, examination of the minimal physics needed to correctly describe metal vs insulator states in both correlated d-electron and s-p electron systems[38,39,63] suggest two major conditions that need to be satisfied: an ideal XC functional should (i) be able to distinguish occupied from unoccupied states, and (ii) have reduced the self-interaction error (SIE).[64,65] Indeed, the wide range of modern XC functional (e.g., local density approximation (LDA)[64], Perdew-Burke-Ernzerhof (PBE)[66], PBEsol[67]) suffer from the SIE arising from spurious interaction of an electron with itself. Since repulsive self-Coulomb interaction exceeds the corresponding attractive self-exchange-correlation, the net SIE is usually positive (repulsive towards electrons), resulting in excessively delocalized wave functions and too high (overly unbound) orbital energies. The tendency to delocalization due to excessive SIE in a given XC potential can be monitored by computing the degree to which the total energy deviates from linearity (the generalized Koopman's condition[65]) as a function of non-integer occupation number. This is a valid "shopping criteria" for selecting an XC functional that can produce spatially compact orbitals with reduces SIE, illustrated in Ref. [32] If this function bows down significantly below the linear (Koopman's) result, we refer to such XC functional as "soft", else it is harder. Meeting the generalized Koopman's condition enforces the minimization of SIE; SIE can lead to underestimation of the band gap energy and even artificially stabilize the non-symmetry-broken configurations. This problem can be fixed by utilizing self-interaction correction (SIC) discussed by Zunger and Freeman[68,69] and by Perdew and Zunger[64], or XC functionals that have reduced self-interaction such as SCAN[70], DFT+U [71-74], or hybrid[75,76] functional. For instance, the SIC applied within DFT for CrO, FeO, CoO, and CuO has been shown to result in band gap opening.[77] A similar tendency has also been demonstrated by DFT+U[38,39] and hybrid functional calculations[78] for 3d oxides. For instance, Fig. 5a illustrates here a case of antiferromagnetic $LaVO_3$ where the choice of a "soft" XC potential (that deviates too much from the linearity condition) can produce a false metal, whereas an XC functional with more reduced SIC gives an insulator (Fig. 5b).

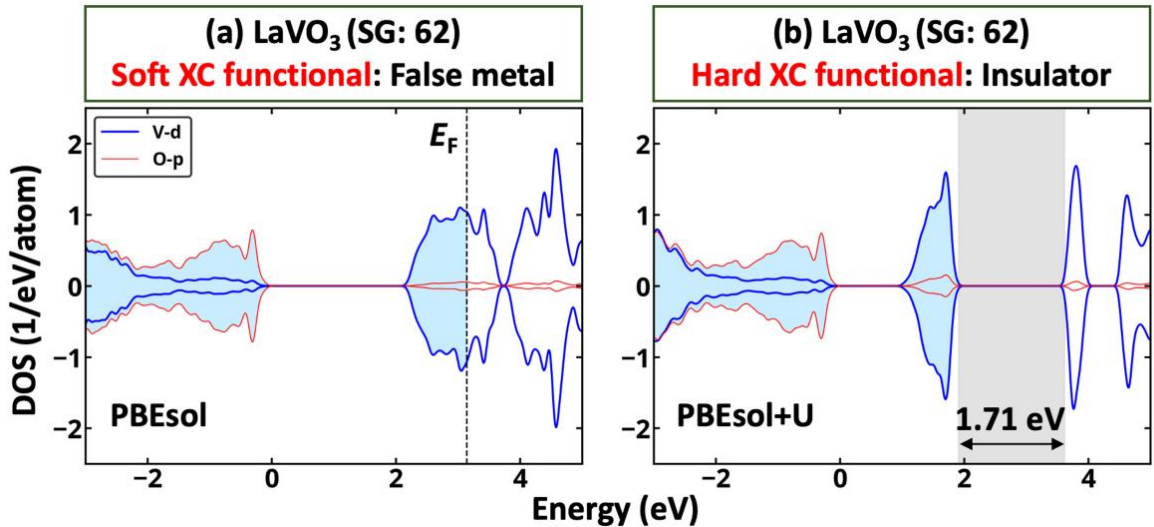

**Figure 5.** (a) Using soft XC PBEsol functional (no U) one predicts a false metal; (b) utilization of hard XC PBEsol+U functional (U =3.5 eV applied on V-d) in band gap opening for experimentally observed AFM-C $LaVO_3$ structure. In both calculations, the lattice vectors and atomic positions are fixed at equilibrium PBEsol values. Occupied states are shadowed in light blue. The band gap is shown in gray. SG denotes the space group number.

**B. Computational need: use a flexible representation for the unit cell allowing symmetry breaking**

One of the computational inputs in performing periodic band calculations is the lattice vectors and internal atomic positions defining the unit cell. Tradition has it that one would use the smallest unit cell, generally as obtained from X-ray diffraction plus Rietveld refinement as, for instance, listed in Inorganic Crystal Structure Database (ICSD)[79]. The latter technique, however, has a large coherence length and thus averages over large volumes, sometimes omitting symmetry-lowering details of the local environment, as seen by experimental techniques with shorter coherence length (x-ray absorption fine structure, pair distribution function, and Raman scattering, to mention a few). Electronic structure methods such as illustrated in Fig. 4 may also be sensitive to local symmetries and magnetic order, but the use of a unit cell with overly restricted size leading to a high average symmetry may not allow this. This problem is well known in alloy theory, where a virtual crystal approximation (VCA) to a substitutional $A_{1-x}B_x$ alloy considers an artificial average structure where each atom is replaced by a virtual <AB> atom, thus artificially raising the symmetry and consequently inhibiting degeneracy removal and local relaxation that are physical effects seen when a larger unit cell is explored.[80] Indeed, there is no general theorem or explanation arguing that the minimal (highest symmetry) unit cell is somehow physically valid.

A simple, common-sense computational test to validate the choice of a unit cell size for a given symmetry is to compute the total energy per atom for a given global cell symmetry but possibly different cell-internal atomic relaxation and spin structures, as a function of (super)cell sizes, and observing if the energy per atom is constant or decreasing. In the latter case, one might find a *polymorphous network*[81], whereby certain local structural features (such as atomic displacements, octahedral tilting), or spin configurations show a *distribution of such local motifs*, rather than a single sharp value. To computationally afford the opportunity of examining if symmetry breaking lowers the total energy, it is important to "nudge" the atomic positions so as to dislodge atoms from possible local minima, and to avoid the practice of wave function symmetrization by equally occupying partner states of a degenerate level (i.e., do not use E (0.5;0.5) for an *e*-level occupied by one electron, but rather E (1;0)).

It turns out that each of these symmetry-breaking modalities can result in energy lowering, leading to local symmetry breaking. The electronic structure can react to the existence of such distribution, even changing from false metal to real insulator, as will be shown below. Indeed, if a system has different local environments (structural or spin) $\{S_i; i=1, N\}$, then its physical property P (band gap, moments, others) cannot be approximated as the properties $<P>=P(S_0)$ of the macroscopically averaged monomorphous structure $S_0$, instead of the correct average $P_{obs}=\Sigma P(S_i)$ of the properties $\{P(S_i)\}$ of the individual, low symmetry microscopic configurations.[38-40,81,82]

Using large supercells and performing minimization with respect to cell internal atomic displacements establishes local displacements. There are two types of displacements in this discussion:

*(a) Local intrinsic displacements:* Those arise from the intrinsic preference of chemical bonding (like bent H-O-H bond angle in water[83], or electronically mandated static Jahn–Teller distortion[84], or due to steric preference such as tilted octahedra[85]). This type of displacements exists even at the lowest temperatures where the in question exists. Intrinsic displacements can be predicted quantum mechanically from the minimization of enthalpy without entropy contributions. They reflect symmetry breaking, such as symmetry lowering off-center atoms. These displacements do not spatially average to zero. As long as there is some length scale of ordering, one will get something finite in the bulk limit.

***(b) Local displacements induced by thermal motion:*** Such displacement represents movements about the low T equilibrium geometry and can be simulated by stochastic movements (i.e., molecular dynamics and Monte-Carlo), causing Urbach tails in absorption. However, if this disorder is truly random, uncorrelated with no form of short or long-range order, then for infinitely large unit cell, the average of such displacements gets zero weight scattering intensity.

The relevance of this discussion on false metals is twofold:

*First,* we will see that degenerate gapped metals tend to become real insulators when intrinsic symmetry breaking removes band degeneracies. Conversely, such systems can stay as metals due to the absence of symmetry breaking. Such absence can occur in two ways: (i) at low-temperature "intrinsic metals" such as $SrVO_3$, where the chemical bonding does not require atomic displacements (as seen in $ABO_3$ compounds with a Goldschmidt tolerance factor[86] near 1, that are stable in ideal, undisplaced cubic phase). This gives real metals even at low T. (ii) Absence of symmetry breaking can also occur at rather high temperatures where strong thermal motions erase the intrinsic displacements. This can give metallic states at high T as in the cases of $BiBaO_3$[87,88] and $SmNiO_3$[89,90]. This illustrates once again the relationship between structural symmetry breaking as a cause of insulating behavior and its absence as the illustration of metallic behavior. Calculations that attempt to simulate the low-temperature phases by ignoring structural symmetry breaking predict false metals for such phases.

*Second*, if one insists on no symmetry breaking, then the XC functional should have an *explicitly* discontinuous (non-differentiable) dependence on the density or density matrix that accounts for spin degeneracies and might predict band gaps. Such discontinuity is missing from all current practical XC approximations. *Thus, using any of the current XC functionals without symmetry breaking polymorphous representation often does not open band gaps*. This paper will illustrate in Sec. III all these modalities of symmetry breaking that transform false metals into real insulators.

## III. Prototype cases of degenerate gapped metals that turn out to be false metals

This section discusses illustrative examples of cases where the omission of *local spin motives* (Subsections A and B), or *local positional motives* (Subsections C, D, E), or spin-orbit coupling (Subsection F) leads to the designation of a compound as a false metal, whereas removal of such artificial constraints reveals they are insulators.

## A. Local spin motifs: Allowing energy lowering FM/AFM spin ordering can convert a false metal to a real insulator

In simplified calculations, compounds are described using nonmagnetic (NM) configurations. However, forcing the NM solution can create artificial stabilization of the false metal case, as illustrated schematically in Fig. 6a. For example, a nonmagnetic scenario for $CuBi_2O_4$ was used in Refs.[91,92] to find a metallic phase having a partially occupied intermediate band in the principal gap, showing an 8-fold band degeneracy at $E_F$ (Fig. 6b). Analogously, the nonmagnetic scenario for NiO (Fig. 6c) obtained by assuming a small unit cell of 1 formula unit (f.u.), gives a p-type false metal.[20] Such result led historically N. Mott to deduce that 3d oxides with an odd number of valence electrons must be metallic in standard band theory.[35] However, experimentally both compounds are insulators. While the above works are only simple illustrations of considering nonmagnetic systems, the nonmagnetic calculations are not rare. For instance, despite groundbreaking studies, recently developed topological databases[20-22] are limited to the NM calculations. In practice, one does not need to guess if a material is NM or magnetic; in most cases, one can use the fact that DFT comes with its intrinsic total energy expression[93] and compare the total energies of magnetic vs NM solutions and pick up the lowest energy one.

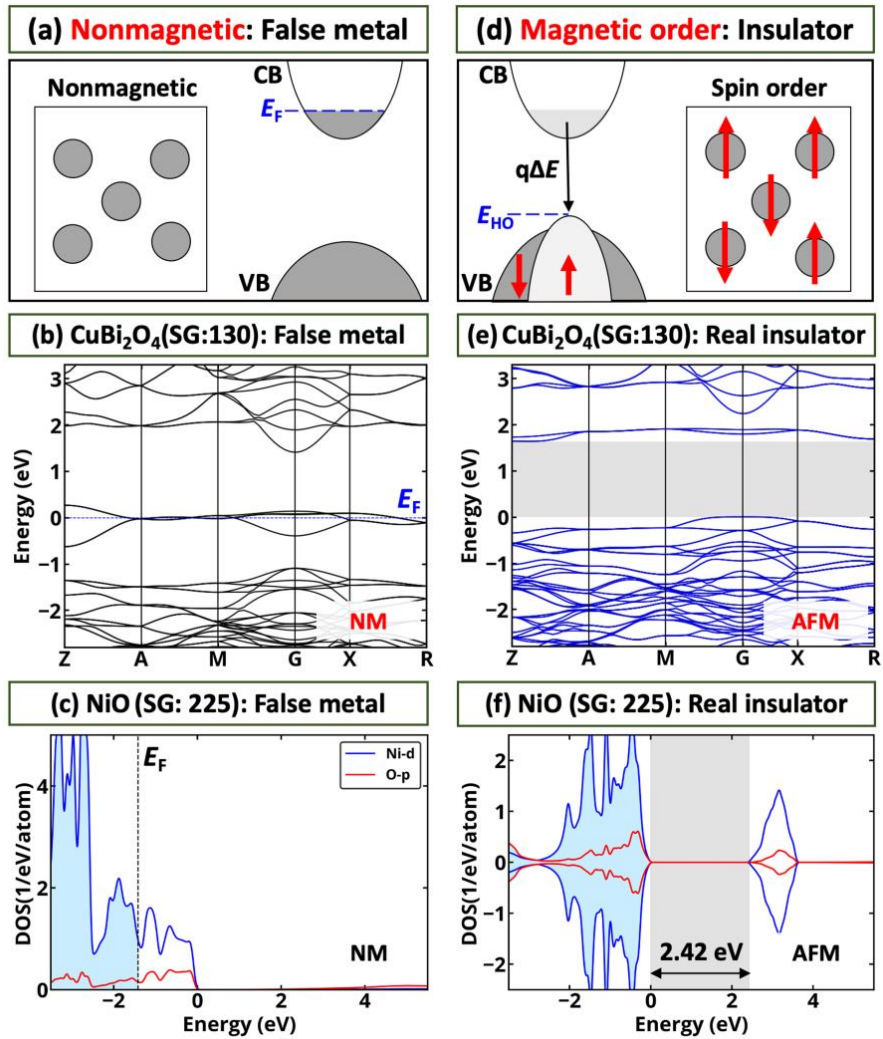

**Figure 6.** Assuming nonmagnetic scenario, (a-c) one predicts a (false) metal; allowing spin-order (d-f) results in the formation of an insulator. (a) Schematic illustration of the electronic structure of a degenerate gapped metal assuming no spin order. (b) Actual calculation for nonmagnetic $CuBi_2O_4$ band structure (using PBE+U with U = 6 eV for Cu-d states) gives a false n-type gapped metal. (c) Actual calculation of nonmagnetic NiO (using SCAN) in a primitive cell containing one formula unit predicting false p-type gapped metal. (d) Schematic illustration of gap opening due to spin order resulting in moving electrons from the conduction band to lower unoccupied orbitals shown by the arrow. (e) and (f) Actual calculations of $CuBi_2O_4$ and NiO allowing AFM magnetic order, showing both systems are insulators. Occupied states are shadowed in light blue. The band gap is shown in gray. SG denotes the space group number. The figures for $CuBi_2O_4$ are redrawn using data from Ref. [31].

*The physical effect that will stabilize an insulating state: magnetic order:* Figure 6d shows schematically how magnetic order can result in band gap opening by creating the empty seats for conducting electrons resulting in the insulator. Such behavior is well-known for AFM compounds where, e.g., doubling the unit cell and allowing magnetic moment formation lead to the band gap opening even in simple band theory.[94] This is indeed the case for both NiO and $CuBi_2O_4$ (Figure 6e,f). Thus, despite the 8-fold band degeneracy at $E_F$ of the *hypothetical* nonmagnetic CuBi$_2$O$_4$[91,92], the lowest total energy magnetically ordered configuration is an insulator with large band gap energy of 1.64 eV according to the PBE+U calculations.[31] Similarly, NiO is the wide band gap insulator with band gap energy of 2.42 eV according to magnetic SCAN calculations. Since the nonmagnetic and AFM structures are identical in atomic positions, the band gap opening in both compounds originates from magnetization only. It should be noted that AFM solutions for both compounds are significantly lower in energy compared to nonmagnetic false metal (i.e., by 0.5 eV/f.u. and by 1.33 eV/f.u. for CuBi$_2$O$_4$[31] and NiO[63], respectively); hence, it is clear that metallic electronic structures are not likely to be realized experimentally in these compounds.

*The experimental situation* is indeed that NiO is the AFM insulator, which has the optical band gap energy of 3.68 eV.[95] This system is often discussed as a Mott insulator where the gapping is caused by electron-electron repulsion.[96] However, as discussed by Zhang et al.[63], the properties of the compound can be described with non-empirical exchange and correlation density-functional (i.e., SCAN) without an on-site interelectronic repulsion, i.e., $U$ = 0 eV. CuBi$_2$O$_4$ is also observed to be a wide band gap magnetic insulator, which recently attracted significant attention for catalysis[97], thus suggesting that 8-band fermions near the Fermi level found in hypothetically nonmagnetic CuBi$_2$O$_4$ are not likely to be realized.

**B. Local spin motifs: Allowing for a polymorphous spin network can convert a false metal to a real paramagnetic insulator**

Paramagnetic (PM) compounds have non-zero local but zero total magnetic moments. Until recently, the properties of such systems have been explored as properties of globally average nonmagnetic structures (Fig. 7a)[14,92,98-101], leading invariably to metallic prediction in contrast with the known insulating properties of many if not most PM ABO$_3$ phases. Because of this, there has been a long-term belief that many properties of PM systems cannot be described within DFT methodology, and higher-order methods (e.g., Dynamical Mean-Field Theory (DMFT) [37,102]) should be applied to get the right result of insulating phase. However, as discussed in Sec. IIB, the properties of a globally average structure should not necessarily be the same as the properties of the system with non-zero local but zero total magnetic moments. To illustrate the limitation of such a naïve PM model equating it with a NM phase, we consider paramagnetic LaTiO$_3$ (SG: 62) and YTiO$_3$ (SG: 62) systems. As shown in Fig. 7b, assuming a NM scenario, both compounds are n-type degenerate gapped metal with 1e/f.u. in the conduction band and large separation between the principal valence and conduction bands, as reported in Refs. [20-22,103,104]. However, experimentally, LaTiO$_3$ and YTiO$_3$ are insulators.[105-108]

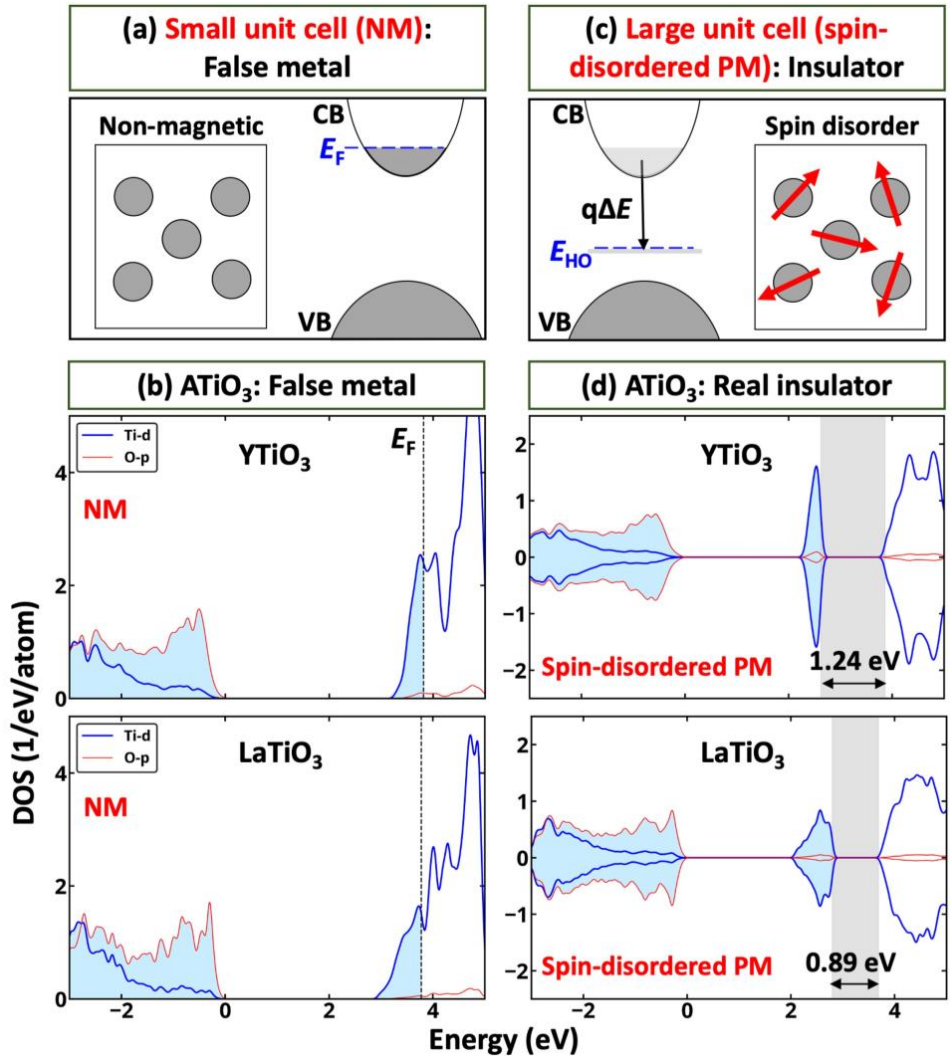

**Figure 7.** When a globally average (nonmagnetic) structure is used with small unit cell, (a-b) one predicts a (false) metal; allowing large unit cell with the spin disorder (c-d) results in the formation of an insulator. (a) Schematic depiction of naïve approximation of the paramagnetic compound using the nonmagnetic configuration resulting in the false metal state. (b) Actual calculation of the density of states for LaTiO$_3$ and YTiO$_3$ within the naïve approximation. (c) Schematic depiction of spin-disordered large unit cell model of a paramagnetic compound resulting in band gap opening due to moving electrons from the conduction band to lower unoccupied orbitals as shown by the arrow. (d) Actual calculation of density of states for LaTiO$_3$ and YTiO$_3$ computed using for 160-atom spin-disordered supercell. All results are presented for PBEsol+U calculations with a U value of 2.5 eV applied on Ti-d states. Occupied states are shadowed in light blue. The band gap is shown in gray.

*The physical effect that will stabilize an insulating state: formation of spin-disordered supercell:* Allowing the atom to have a non-zero magnetic moment on each site can result in energy lowering and band gap opening (Fig. 7c) due to moving the conduction electrons to the lower energy level. It has been shown that modeling the paramagnetic compound as a spin-disordered system (each atom has non-zero spin, but the spins in the system are disordered) can be used to effectively reproduce experimental properties of binary[38,63] and ternary[39,40] paramagnetic compounds within the DFT calculations. These works thus demonstrated that it is possible to model paramagnetic systems with DFT calculations despite the counter, long-term beliefs. Application of this model to $LaTiO_3$(SG: 62) and $YTiO_3$(SG: 62) results in insulators (Fig. 7d) with PBEsol+U band gap energies of 0.89 and 1.24 eV, respectively. As will be discussed in Sec. IV, the formation of an insulator is accompanied by electron localization on Ti atoms, resulting in e-trapped states.

*The experimental situation:* Paramagnetic $LaTiO_3$(SG: 62) and $YTiO_3$(SG: 62) are insulators as reported by multiple studies and confirmed by both temperature dependence of resistivity and photoemission data.[105-108] These results thus prove that prediction of metals for both $LaTiO_3$ and $YTiO_3$ is the artifact of the nonmagnetic PM model, while the spin-disordered approximation can successfully represent the main experimental predictions for the PM phases, which is in line with that reported by Varignon *et al*.[39,40]

*In which type of compounds would this insulator stabilization by spin-disordered supercell occur?* Since the model of spin-disordered systems is relatively new, the set of compounds where the band gap opening has been reported is currently limited to binary MnO, FeO, CoO, and NiO[38,63] and set of 3d $ABO_3$ oxides[39,40]. Specifically, it has been demonstrated that the consistent way to calculate the properties of PM phase as the polymorphous statistical average over the ensemble of microscopic configurations and not the property of macroscopically average structure (see Sec. IIB).

We close this subsection by discussing the manner in which our description of a PM phase constitutes a generalization of the well-known low-temperature spin ordered phase. The low-temperature AFM phase generally inherits some of the properties of the PM phase. The simple illustration of such behavior is the comparison of electronic properties of AFM and PM phases of $YNiO_3$(SG: 14), showing that both phases have similar band gap energies of 0.59 and 0.49 eV (Fig. 8), respectively according to PBEsol+U calculations. In contrast, the NM $YNiO_3$(SG: 62) phase is a degenerate gapped conductor with a partially occupied intermediate gap state, which has higher total energy, about 0.13 eV/f.u. above both AFM and spin-disordered phase. Indeed, the low-temperature AFM phase can have spin and positional local motifs that are a subset of those in the high-temperature phase. For example, the AFM phase can have a single spin motif – each spin-up is coordinated by all spin-down sites, whereas the PM phase represents a generalization, having a distribution of local motifs where each spin-up is locally coordinated by $m$ spin-up and $N-m$ spin-down, where N is the coordination number and $0<m<N$. Similarly, the high-temperature phase can have a *distribution of local displacements* resembling geometrically the global structural motif that is responsible for long-range order in the low-temperature ground state. An example is the high-temperature trigonal phase of FeSe having locally orthorhombic distortions mimicking the globally orthorhombic stable low-temperature phase.[82] In general, the transition from high-temperature PM to the low-temperature ordered phase involves ordering vectors that select from the PM phase certain spin and structural motifs that are stabilized.

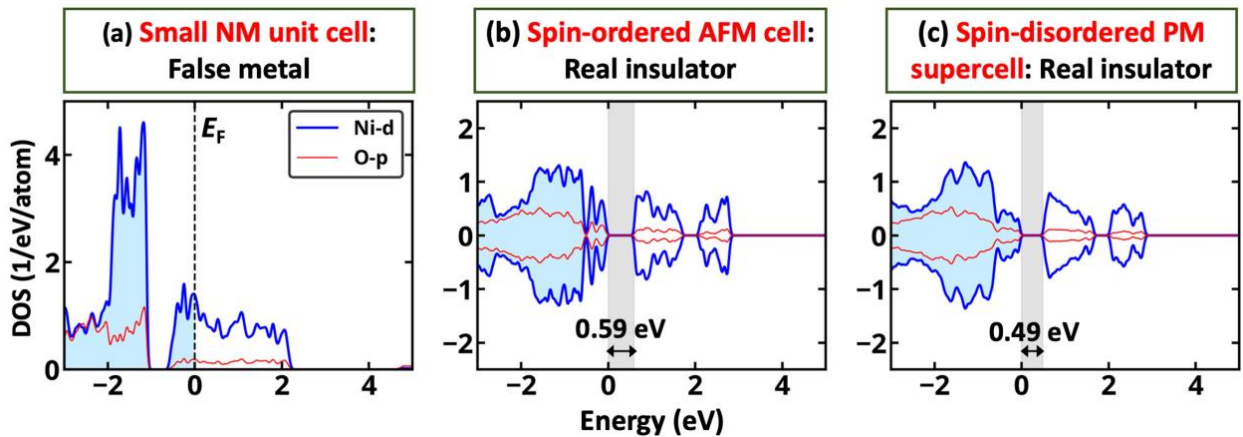

**Figure 8.** Comparison of electronic structures for (a) naïve nonmagnetic, (b) antiferromagnetic, and (c) 160-atom spin-disordered paramagnetic phases of $YNiO_3$. The results demonstrate that electronic structure of the spin-disordered paramagnetic phase of $YNiO_3$ is closer to antiferromagnetic order than to the naïve nonmagnetic approximation. Results are presented for PBEsol+U exchange correlation functional with a U value of 2 eV applied to Ni-d states. Occupied states are shadowed in light blue. The band gap is shown in gray.

**C. Local structural motifs: Enabling energy-lowering bond disproportionation can convert a false metal into a real insulator**

In traditional first-principles calculations, compounds are usually described with the smallest possible primitive cells where each species is often represented via a so-called Single Local Environment (SLE), as shown schematically in Fig. 9a. While this approach allows a proper description of the properties of some compounds, it does not guaranty that the formation of structurally inequivalent Wyckoff positions does not lower the energy of such structures. This can lead, for instance, to a Double Local Environment (DLE) where the chemical bonding can mandate a larger unit cell. Examples of the consequence of assuming SLE behavior are shown in Fig. 9 for monoclinic $TiO_2$ (space group 12, a structure often denoted as "(B)"[109]), doped with Li interstitial and for cubic SrBiO$_3$. The consequences of this assumption are that the former case has the $E_F$ in the conduction band with 1e/f.u., whereas the latter case has the $E_F$ in the valence band with 1h/f.u., as shown in DFT calculations in Fig. 9b and 9c, respectively. However, experimentally, both systems are insulators.[110,111] There are a number of cases when using such approximation invalidates the proposed new functionality, e.g., Nanda et al.[112,113] suggested cubic SLE SrBiO$_3$ as a topological compound, but this structure is not the stable phase.

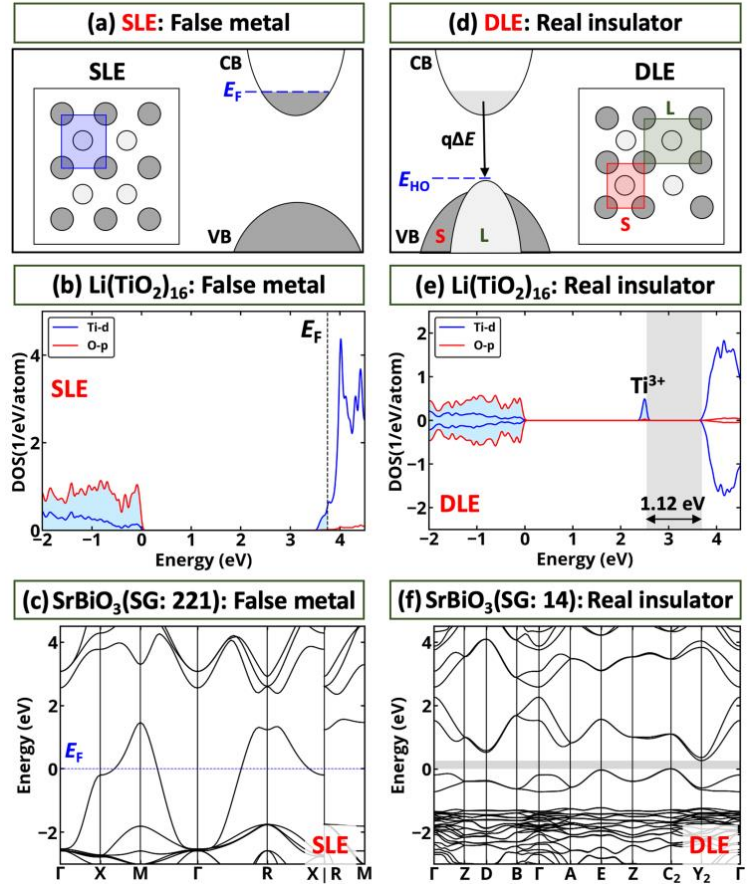

**Figure 9.** Assuming a single local environment (SLE), (a-c) one predicts a (false) metal; allowing double local environment (DLE) (b-f) results in the formation of an insulator. (a) Schematic illustration of degenerate gapped metal having a single local environment. (b) HSE calculation of density of states for nonmagnetic SLE 16 f.u. supercell of monoclinic TiO$_2$ (SG: 12) containing Li interstitial atom demonstrating that it is n-type degenerate gapped false conductor with Fermi level in the conduction band. (c) PBE+SOC calculation for SLE cubic SrBiO$_3$ (SG: 221) showing that it is p-type degenerate gapped false conductor with Fermi level in the valence band. (d) Schematic illustration of degenerate gapped metal having double local environment resulting in band gap opening due to moving electrons from the conduction band to lower unoccupied orbitals as shown by the arrow. (e) HSE density of states for 16 f.u. supercell of monoclinic TiO$_2$ (SG: 12) containing single Li interstitial showing that the system is an insulator with e-trapped intermediate band caused by localization of electron on the part of Ti sublattice – formation of DLE. Occupied states are shadowed in light blue. The figure is redrawn using data from Ref. [114]. (f) PBE+SOC band structure for DLE monoclinic SrBiO$_3$ (SG: 14) showing that the compound is an insulator with band gap energy (shown in gray) of 0.26 eV. SG denotes the space group number.

*The physical effect that will stabilize an insulating state is creation of different local environments:* Fig. 9d illustrates how the formation of different local environments in a degenerate gapped metal can result in band gap opening and total energy lowering. This turns to be the case for both $TiO_2$:Li and cubic $SrBiO_3$, which both spontaneously disproportionate to lower energy structures when symmetry breaking is allowed. The false metal Li-doped monoclinic $TiO_2$ reconstructs to lower energy insulator (with a HSE band gap energy of 1.12 eV) having an e-trapped intermediate band (Fig. 9e). As will be discussed in Sec. IV, the formation of the e-trapped state and the band gap opening are due to the ability for Ti atoms to change the formal oxidation state from $Ti^{4+}$ to $Ti^{3+}$. This is evidenced by concomitant structural changes where clearly district local environments are formed - the average $Ti^{4+}$-O bond length is 2.02 Å, while the corresponding value for the $Ti^{3+}$-O bond is 2.09 Å. The same tendency is observed in $SrBiO_3$(SG: 221) supercell that spontaneously disproportionates to monoclinic $SrBiO_3$(SG: 14)[111], where Bi has a double local environment (DLE), lowering the system energy by 0.14 eV/atom. The resulting system is an insulator with the PBE+SOC band gap energy of 0.26 eV having h-trapped intermediate band (Fig. 9f). This reaction is caused by the ability of Bi to disproportionate to $Bi^{3+}$ and $Bi^{5+}$,[115-117] which is confirmed by structural analysis; monoclinic $SrBiO_3$ (SG: 14) contains structurally different Bi atoms with the average Bi-O bond lengths of 2.17 and 2.34 Å. One should note, however, that herein FOS is used as a label without assigning any physical meaning. Indeed, FOS does not describe the charge density distribution in solids due to "self-regulating response"[118] - a change of cation charge is usually counteracted by opposing change on the ligand.

*The experimental situation* is that the formation of e-trapped states originated from reduced Ti atom in $TiO_2$:Li systems are well-known phenomena discussed by different groups[110,114,119-121] as the cause for low electronic conductivity of $TiO_2$:Li systems formed during lithiation of $TiO_2$-based electrode. Experimental investigations also confirm that despite numerous studies reported on the cubic-like structure of $SrBiO_3$ (SG: 221)[111,116], it always exists in monoclinic structure (SG: 14). Moreover, as indicated above, the compound is always found to be an insulator, and the formation of structurally different Bi atoms has been often referenced as $Bi^{3+}$ and $Bi^{5+}$.[116] These results thus prove that cubic $SrBiO_3$ is the example of false metal caused by forcing the SLE in the compound that exhibits spontaneous energy lowering by bond disproportionation and formation of the DLE phase.

*In which type of compounds would this insulator stabilization occur?* The DLE behavior for compounds being a degenerate gapped ***n-type metal in the precursor SLE state*** is common for systems having cations which can exist in different oxidation states (e.g., $Ti^{4+}/Ti^{3+}$, $Ce^{4+}/Ce^{3+}$) where the splitting between the orbital energies of the two FOS is large. $TiO_{2-x}$, $CeO_{2-x}$, and $V_2O_{5-x}$ are well-known examples of insulators having different local environments and e-trapped in-gap states localized on the reduced metal ions.[120,122,123] This set of examples can be further extended to a wide family of ternary early transition metal oxides where the sum of FOS differs from 0 (e.g., $TiO_2$:Li).[110,114,119] The DLE behavior for compounds being a degenerate gapped ***p-type metal in the precursor SLE state*** can be found either in (i) nonmagnetic insulators that include high-Z elements (e.g., $BaBiO_3$[87,115,124], $SrBiO_3$[117], $CsTlF_3$[115], $CsAuCl_3$[115], and $CsTe_2O_6$[115]), or in (ii) compounds having band gap opening as the result of the superposition of disproportionation and magnetism. The classic examples of such materials are rare-earth nickelates ($ANiO_3$ with A = Sm, Eu, Y, and Lu[39,115,125]) and $CaFeO_3$[115]. It is interesting that disproportionation in d electron compounds such as $SmNiO_3$ was initially described as an effect enabled specifically by electron correlation.[126] However, the disproportionated structure was, in fact, also obtained in a straight DFT calculation based on broken symmetry[125], reflecting the physics of broken symmetry as indicated here.

## D. Local structural motifs: Enabling energy lowering pseudo-Jahn-Teller-like distortions can convert a false metal into a real insulator

The degeneracy-removing electronically-induced Jahn-Teller $Q^-_2$ distortion as well as the structurally induced (e.g., by steric effects) pseudo-Jahn-Teller $Q^+_2$ distortions are known to exist in numerous perovskite compounds[127], but were often ignored when in describing metallic or insulating behavior. Simple approximations often assume high symmetry cubic structures that cannot geometrically accommodate such local symmetry lowering distortions. This assumed cubic structure is especially popular in the description of the Mott insulators, where it has been widely used for DMFT Hamiltonian mapping[128] in the potential degenerate gapped metals at least until recently. For the case of potential degenerate gapped metals, forcing the cubic symmetry can artificially move $E_F$ to CB or VB (Fig. 10a). $LaMnO_3$ is an example of such behavior: It is a metal with $E_F$ in CB and 1e/f.u. free carriers if its pseudo-Jahn–Teller distortions are ignored (Fig. 10b). These results conflict with available experimental data showing that the compound is AFM insulator with a wide band gap at low temperature.[129]

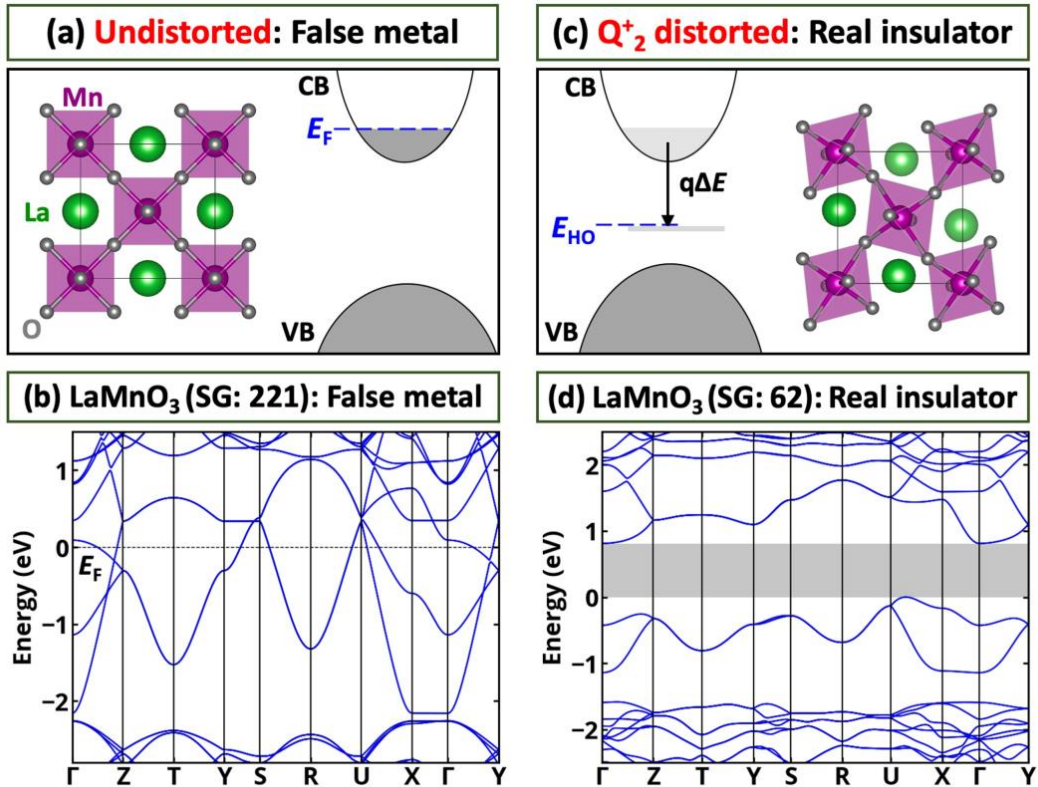

**Figure 10.** Ignoring the pseudo-Jahn Teller $Q^+_2$ distortion, (a,b) one predicts for LaMnO$_3$ a metal; (c,d) allowing energy lowering $Q^+_2$ distortion results in the formation of insulator. (a) Schematic illustration of ideal cubic ABO$_3$ structure having all octahedra identical with equivalent Mn-O bond length for each metal with Fermi level in conduction band. (b) Actual calculation (using the SCAN XC) for cubic AFM LaMnO$_3$ demonstrating that it is n-type degenerate gapped false conductor with Fermi level in the conduction band. (c) Schematic illustration of $Q^+_2$ symmetry breaking in cubic ABO$_3$ structure resulting in inequivalent B-O bond lengths and band gap opening due to moving electrons from the conduction band to lower unoccupied orbitals as shown by the arrow. (d) Actual calculation (using the SCAN XC) of band gap opening in AFM LaMnO$_3$ as a result of $Q^+_2$ distortion. The band gap is shown in gray. SG denotes the space group number. The figure is redrawn using data from Ref. [130].

*The physical effect that will stabilize an insulating state is $Q^+_2$ distortions:* The deviation from 1 of the Goldschmidt tolerance factor can lead to different energy lowering reconstruction resulting in a change of electronic properties. For the case of LaMnO$_3$, the compound exhibits spontaneous $Q^+_2$ distortions[127,130,131] resulting in band gap opening (see Fig. 10c,d). Specifically, while cubic LaMnO$_3$ has 0 rotation angle and all Mn-O bonds of equivalent to each other, in the distorted orthorhombic structure there is octahedra rotation and non-equivalent Mn-O bonds – in a cubic structure, Mn-O bond length is 1.94 Å, while for orthorhombic one, Mn-O bond lengths are 1.91, 1.97, and 2.21 Å. According to the SCAN calculations, LaMnO$_3$ is an insulator having e-trapped intermediate band occupied by 1e/f.u. and band gap energy of 0.81 eV. Importantly, this band gap opening is observed without using any electron-electron repulsion U value, suggesting that the true origin of the band gap opening is energy lowering structural symmetry breaking. The detailed analysis of such symmetry breaking and its effect on the electronic structure has been recently documented by Wang *et al.*[130], who also noted a significant change of effective masses in the compound as the result of symmetry lowering.

*The experimental situation:* Orthorhombic LaMnO$_3$ is an AFM-A insulator having clearly distortions with respect to ideal cubic form.[129,132] According to previous experimental data[129,132], the compound is expected to be cubic at high temperature, which indeed should be a metallic system with one of the highest known concentration of free carriers for degenerate gapped metals(e.g., Fig. 10b). However, it has been shown that cubic LaMnO$_3$ is unrealizable as despite having cubic-like structure at high temperature, the MnO$_6$ octahedra are still noticeably distorted locally as demonstrated by Qiu et al.[133] It should be noted that at high temperature the LaMnO$_3$ can have the metallic-like behavior based on the analysis of resistivity vs temperature.[134] This phenomenon is still not completely understood yet. One should note that while LaMnO$_3$ discussed herein is the example of $Q^+_2$ distortion resulting in band gap opening, there is the range of other possible octahedra distortion mechanisms discussed in the literature as the cause of band gap opening. For instance, LaTiO$_3$[135], KCrF$_3$[136], and KCuF$_3$[137] exhibit the different $Q_2$ distortions.

## E. Local structural motifs: Allowing for spontaneous defect formation can convert a false metal into a real insulator

The formation of nonstoichiometric compounds is usually attributed to a growth effect rather than to a thermodynamically mandated specific instability. Hence, spontaneously defected compounds are usually neglected in many theoretical and experimental studies. This can lead to an incorrect prediction of a degenerate gapped metal. Fig. 11a,b show calculations for stoichiometric defect-free Ba$_4$As$_3$ and Ag$_3$Al$_{22}$O$_{34}$ having a large internal gap and $E_F$ in principal valence and conduction bands, respectively. However, these compounds have never been realized experimentally under the stoichiometric conditions. All attempts to synthesize metallic $Ba_4As_3$ result in the formation of nonstoichiometric $Ba_4As_{3-x}$ having insulating properties.[138] The same tendency is also found in attempts to synthesize $Ag_3Al_{22}O_{34}$[23]. Although both these compounds have not been widely studied theoretically, they attracted some attention. Specifically, stoichiometric $Ag_3Al_{22}O_{34}$ has been predicted as a potential intrinsic transparent conductor[24], while stoichiometric $Ba_4As_3$ is shown in the Materials Project database[43]. Moreover, $Ba_4As_3$ represents a wide family of potential false metals (e.g., $Ba_4Bi_3$, $Sr_4Bi_3$, and $Yb_4Sb_3$) recently predicted to have topological properties in $Ba_4As_3$-like structure.[20-22]

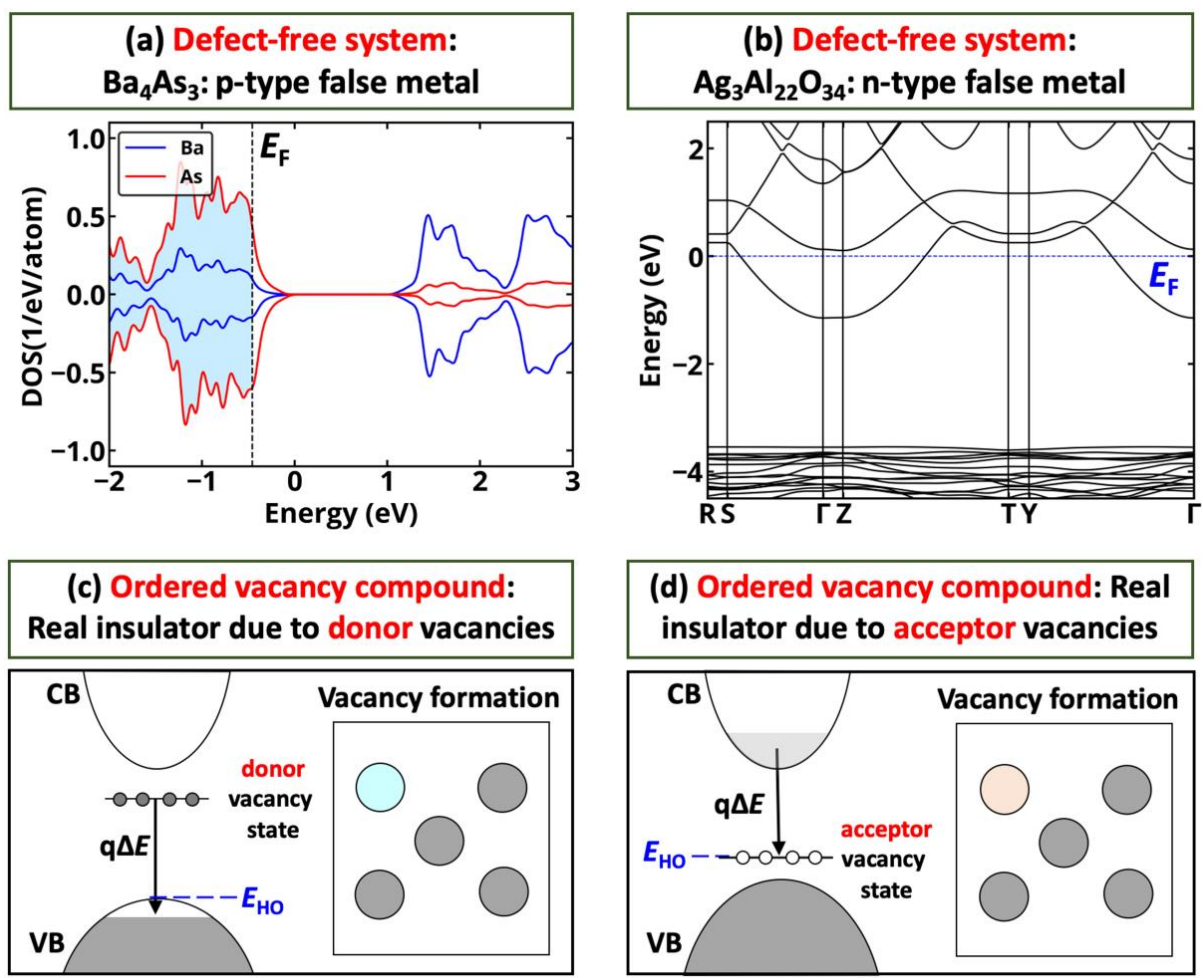

**Figure 11.** When defect formation is disallowed, (a-b) one predicts a (false) metal; allowing vacancy formation (c,d) leads to spontaneous nonstoichiometry and formation of the real insulator. (a) Density of states for $Ba_4As_3$ showing that it is potential p-type degenerate gapped metal according to PBE calculations. Occupied states are shadowed in light blue. (b) Band structure of $Ag_3Al_{22}O_{34}$ showing that it is potential n-type degenerate gapped metal as computed using PBE+U with U value of 5 eV applied for Ag-d states. (c,d) Schematic illustrations of the formation of donor/acceptor vacancy in p-/n-type degenerate gapped metals resulting in electron-hole recombination reducing the vacancy formation energy. Donor and acceptor vacancies are shown as light blue and beige colors, respectively.

*The physical effect that stabilizes an insulating state: spontaneous vacancy formation:* For some degenerate gapped metals, the formation of point defects (e.g., intrinsic vacancies) can be spontaneous due to an electronic instability.[23,31] The point is that while in traditional insulators, vacancy formation requires endothermic breaking the bonds, for degenerate gapped conductors, the presence of carriers in conduction or valence band can change the defect physics, leading to *spontaneous* defect formation. For instance, for p-type degenerate gapped conductors (Fig. 11c), the formation of donor vacancy results in the moving electrons from the donor level to the hole states in the valence band, which can restore the part of the energy needed to form the vacancy. Similarly, for n-type degenerate gapped metal (Fig. 11d), the formation of acceptor vacancy can result in decay of conducting electrons to the acceptor level restoring part of the energy needed to create the vacancy. Such electron-hole recombination can result in spontaneous vacancy formation, which can induce significant deviation from stoichiometry at low-temperatures.[23,31,139,140] To examine the possibility of the instability of stoichiometric $Ba_4As_3$ and $Ag_3Al_{22}O_{34}$, we study the formation of As vacancy (donor) in $Ba_4As_3$ and Ag vacancy (acceptor) in $Ag_3Al_{22}O_{34}$. Taking into account all experimentally known stoichiometric phases in Ba-As and Ag-Al-O phases, the range of chemical potentials for stability of $Ba_4As_3$ and $Ag_3Al_{22}O_{34}$ is predicted by calculations of energy convex hull[141] which defines the ACS having energy lower than any linear combination of any competing phases at the corresponding compositions. Thus, we find that $Ba_4As_3$ is on the convex hull while $Ag_3Al_{22}O_{34}$ is slightly above the convex hull (Fig. 12a,c) – i.e., there are no chemical potentials at which it is thermodynamically stable.

The computed vacancy formation energies (Fig. 12b,d) demonstrate that stoichiometric $Ba_4As_3$ and $Ag_3Al_{22}O_{34}$ are unstable with respect to the spontaneous formation of donor and acceptor vacancies, respectively. Moreover, the vacancy formation results in the reduction of carrier density due to electron-hole compensation – each As vacancy in $Ba_4As_3$ removes 3h from the valence band, and each Ag vacancy in $Ag_3Al_{22}O_{34}$ depletes 1e from the conduction band. These results suggest that both compounds are unstable with respect to vacancy formation and hence unlikely to exist in stoichiometric forms (a far larger concentration of defects than considered here might exist).

*The experimental situation*: The perfectly stoichiometric $Ba_4As_3$(SG: 220) has never been synthesized. Stoichiometric $Ba_4As_3$ is a compound inspired by experimentally reported nonstoichiometric $Ba_4As_{3-x}$(SG: 220), which is indeed a simple insulator.[138] $Ba_4As_{3-x}$(SG: 220) and $Ba_4As_3$(SG: 220) structures are similar, with the only difference being that each As site has partial occupancy and that the As vacancies are randomly distributed across the sample. While $Ag_3Al_{22}O_{34}$ structure has been suggested experimentally[142], all attempts to synthesize the compound lead to close stoichiometry $Ag_{3-x}Al_{22}O_{34+y}$ having similar structure and insulating properties.[23] Taking into account the above defect calculations, we conclude that the experimentally observed nonstoichiometry of the compounds is caused not by the growth effect but by the Fermi level instability: the energy lowering caused by electron-hole recombination as the result of vacancy formation is larger than energy gain needed to break the chemical bonds.

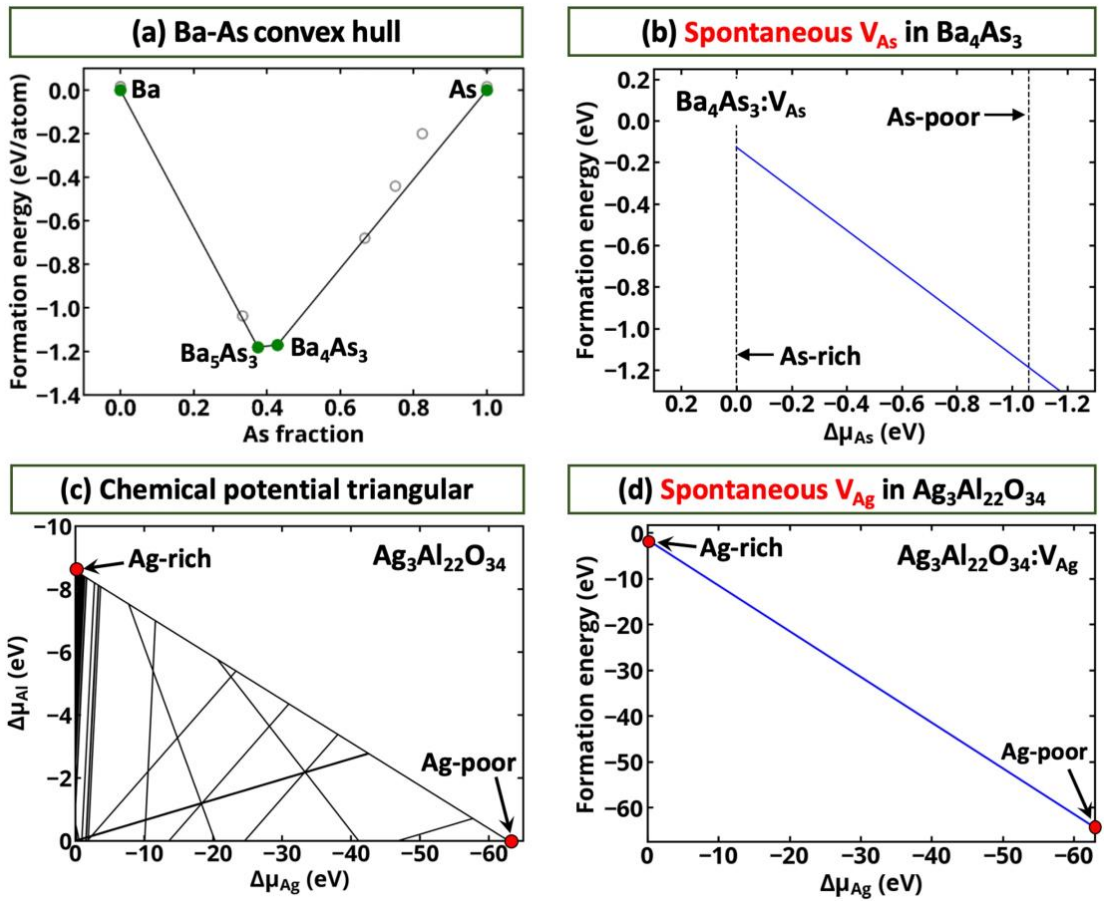

**Figure 12.** Spontaneous formation of As vacancy in $Ba_4As_3$ and Ag vacancy in $Ag_3Al_{22}O_{34}$. (a) Ba-As energy convex hull computed using all experimentally known stoichiometric Ba-As compounds. The compounds above the convex hull are shown by the empty gray circles. (b) Defect formation energy for As vacancy in $Ba_4As_3$ as the function of As chemical potentials. (c) Stability triangular for $Ag_3Al_{22}O_{34}$ demonstrating that the compound is unstable with respect to decomposition to compering phases when known experimental Ag-Al-O compounds are taken into account. (d) Defect formation energy for Ag vacancy in $Ag_3Al_{22}O_{34}$ as the function of Ag chemical potentials. The figure for $Ag_3Al_{22}O_{34}$ is redrawn using data from Ref. [23]. The results for $Ag_3Al_{22}O_{34}$ are given for PBE+U calculations with a U value of 5 eV applied on Ag-d states. For $Ba_4As_3$, the results are given for PBE calculations. Fitted Elemental-phase Reference Energies (FERE)[143] is used to correct the elemental chemical potentials.

**F. Spin-Orbital motifs: Spin-orbit coupling in high-Z compounds can convert false metals to insulators**

A more specialized effect, pertaining to compounds containing a high-Z atom is the effect of gapping a false metal due to spin-orbit coupling. In traditional first-principles studies[43-45], the investigation of materials properties is usually limited to the non-SOC calculations. This computational approximation works sufficiently well for light elements, and the large set of material properties can be described well within this simplified approximation.[144] In general, the SOC effect on electronic structure is usually discussed in terms of spin-orbit splitting. For degenerate gapped metals, SOC can result in the splitting of a conduction or valence band, giving a true insulator. An example of such behavior is $CaIrO_3$ (SG: 63) – a compound which without SOC is predicted to be p-type conductor[145] with the Fermi level in the principal valence band (Fig. 13a) even when a hard XC functional (e.g., HSE) is used. However, this result contradicts experimental data suggesting that $CaIrO_3$ is an insulator based on the measurement of electronic resistance vs temperature.[146] When SOC is applied, the system becomes an insulator with band gap energy (see Fig. 13b)[145], which is due to the splitting of the valence band. A similar example of band gap opening has been reported for $Sr_2IrO_4$.[147] These explorations thus demonstrate the important contribution of the SOC in the false metal - real insulator transition, suggesting that SOC should be taken into account for the prediction of real degenerate gapped metals. It should be also be noted that the band gap opening is only observed with XC functional satisfying Sec. IIA XC conditions – application of SOC on top of PBE functional often does not result in band gap opening. It should be noted that recently developed topological databases[20-22] provided the results for electronic structures of many compounds with included SOC. However, since all calculations were performed with "soft" XC functionals, it is likely that databases have a number of false predictions. For instance, both $Sr_2IrO_4$ and $CaIrO_3$ are predicted to be topological metals.[20-22]

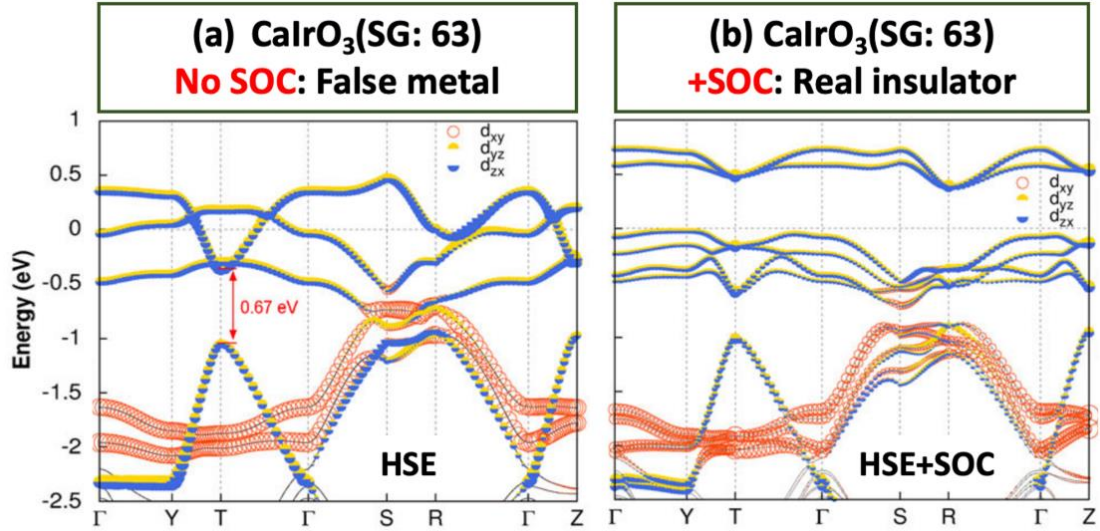

**Figure 13.** When SOC is ignored, (a) one predicts a (false) metal; adding SOC (b) results in the formation of an insulator. Band structure projected onto $d_{xy}$, $d_{yz}$, and $d_{zx}$ orbitals for $CaIrO_3$ computed with (a) HSE and (b) HSE+SOC. The highest occupied level is at 0 eV. The radii of semicircles and circles are proportional to the weights of the orbitals. SG denotes the space group number. The figure is reproduced with permission from Kim et al. Phys. Rev. Lett. 115, 096401 (2015), Copyright (2015) by the American Physical Society.

## IV. Symmetry breaking in degenerate gapped metal leading to localized trapped carrier states

We have seen that the symmetry-breaking modes discussed herein can (i) result in band gap opening without the formation of any in-gap states, as demonstrated in Fig. 6f for AFM NiO or (ii) cause splitting of a subband from the continuum (Fig. 1a), forming an intermediate band within the principle band gap (Fig. 1b). Inspection of the wave function of such split-off bands created via symmetry breaking can reveal the nature of in-gap states and its correlation to local structural distortions. The orbital character of the split-off intermediate band can either (a) mimic that of the nearest band edge from which it was split, corresponding to a normal situation of degeneracy removal with the possibly trapped carrier (because the intermediate band is not connected via dispersion to any other band), or (b) the split-off band can show localization of certain sublattices (see below $TiO_2$:Li) behaving as a defect level, except that the defect here is *electronic*. This situation is common when the localizing sublattice atom can exist in more than one FOS, such as $Ti^{3+}$ and $Ti^{4+}$. In the dilute defect limit, this latter situation is often discussed as polaron – a quasiparticle originating from the interactions of electrons/holes with a lattice ion, often causing local distortions.[148,149] Inspection of the wavefunctions (Figure 14) of the broken symmetry cases IIIA-III-F reveals the following instances of intermediate bands:

**Polaron-like states in $TiO_2$:Li:** Doping of ordinary insulators can result in different compensation mechanisms that remove free carriers, including the formation of electronic defects being polarons. A polaron can form without doping or with doping. An example of the latter is the widely discussed e-doped $TiO_2$ systems[114,122,150-152] and is demonstrated on the example of density of states for Li-doped $TiO_2$ shown in Fig. 9e. Indeed, Li doping of $TiO_2$(SG: 12) results in disproportionation and carrier localization – one Li converts one $Ti^{4+}$ to $Ti^{3+}$. This is in agreement with the fact that the calculated charge density corresponding to the in-gap states shows localization of e-trapped intermediate band on single Ti atom (see Fig. 14a). Indeed, the existence of Li induced $Ti^{3+}$ states have been used to quantify the Li content coming from surface reaction.[110] Among the considered compounds, $TiO_2$:Li has the narrowest band of in-gap states, which is mainly due to a low concentration of reduced $Ti^{3+}$ weakly interacting in the system.

Conventional split-off bands due to symmetry breaking can be divided into electron traps or hole traps:
**Formation of electron-trapped valence band maxima:** DFT depiction of early transition metal oxides (i.e., $LaVO_3$, $YTiO_3$, $LaTiO_3$) exhibit the formation of an *occupied* "upper Hubbard Bands" that contain trapped electrons (see Fig. 5b, 7d, 10d). $LaVO_3$, $YTiO_3$, $LaTiO_3$, and $LaMnO_3$, each having a structurally unique 3d metal atom with roughly the same wavefunction amplitude for the trapped carriers (see Fig. 14b-d), suggesting that the compounds can be considered as $La^{3+}V^{3+}O_3^{2-}$, $Y^{3+}Ti^{3+}O_3^{2-}$, $La^{3+}Ti^{3+}O_3^{2-}$, and $La^{3+}Mn^{3+}O_3^{2-}$. The e-trapped states in $ABO_3$ compounds are reported in photoemission studies showing the existence of in-gap occupied states located about 1.3-1.5 eV below the principal conduction band (or the Fermi level).[105-108] Moreover, for $YTiO_3$, the in-gap state was attributed to the reduced $Ti^{3+}$, which is in line with the above discussion.

**Formation of hole-trapped conduction band minima:** DFT depiction of $YNiO_3$, $SrBiO_3$, and $CuBi_2O_4$ shows clearly distinct formation of h-trapped states in Fig. 6e, 8b, 9f. Here, $YNiO_3$ and $SrBiO_3$ are the compounds having disproportionation in the low-temperature phase, where one can expect different charge density distribution on structurally different B atoms. For AFM monoclinic $YNiO_3$, there are two structurally different Ni sublattices with one magnetic and one nonmagnetic. These Ni sublattices have different contributions to electronic structures. This is shown in Fig. 14e, demonstrating that the h-trapped state is not only localized on O but also has contribution from Ni atoms. Here, due to DLE structure, one type of Ni atoms contributes significantly more than another one. This picture is similar to that for nonmagnetic $SrBiO_3$ shown in Fig. 14f.

The formation of hole trapped conduction band minima and electron trapped valence band maxima is often characteristic of existence of electronically distinct metal ions capable of carrying different valences. Such "electronic defects" are analogous to "atomic impurities/defects". The latter traditional defect physics described by Koster and Slater[153] and extended to 3D semiconductors by Hjalmarson et al.[154] depict the formation of in-gap states in the language of the perturbation caused by impurity **I** substituting a host atom H, i.e., $\Delta V_{imp}$=V(I)-V(H). If the magnitude of this perturbation exceeds some threshold value (that depends on the band width), in gap split-off states can form. Electronic defects are an analogous problem to atomic defects in that the perturbation can be defined as the difference of potential for a real insulator having in-gap states and corresponding false metal, for instance, as $\Delta V_{elec}$= V($Ti^{3+}$)- V($Ti^{4+}$).

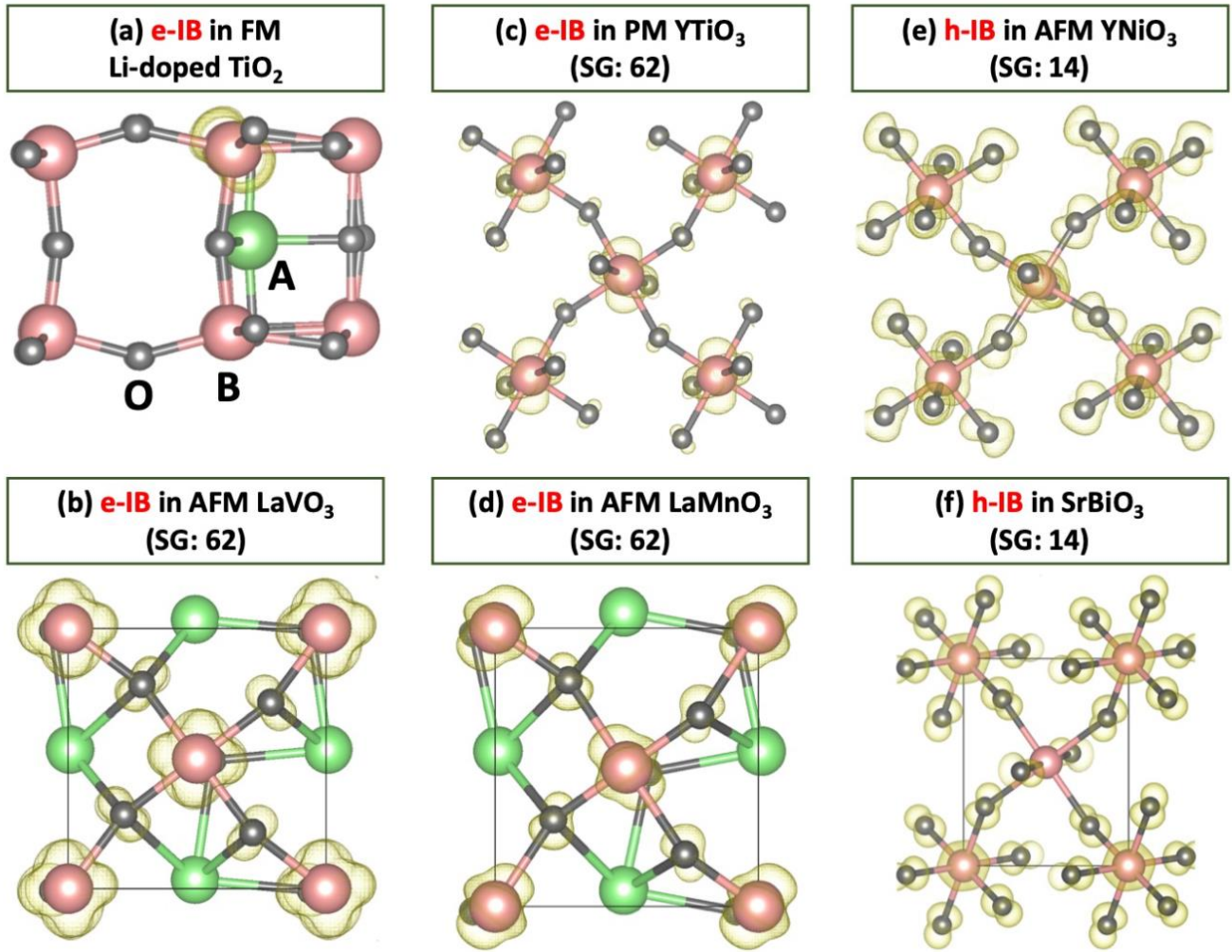

**Figure 14.** Wavefunction amplitudes (yellow isosurfaces) computed for electron (e) or hole (h) trapped intermediate bands (IBs) in TiO$_2$:Li and ABO$_3$ compounds demonstrating carrier-trapping. For all cases, the isosurface is set at 0.01 e/Bohr$^3$. A, B, and oxygen atoms are shown as green, pink, and gray, respectively. SG denotes the space group number.

***New doping physics ("antidoping") in real insulators originated from false metals:*** While the above examples are pristine false metals, recent discussion has focused on doping them.[120,155,156] Specifically, it has been discovered that n-type doping of compounds having h-trapped IBs (e.g., YNiO$_3$[120] and SmNiO$_3$[155,156]) can result in a shift of intensity of intermediate band towards the *valence band* leading to reduction of conductivity and band gap opening. The similar band gap opening mechanisms can also be observed for hole doping the compounds having e-trapped intermediate bands[120], where adding holes results in band shift towards the conduction band with following band gap opening. Such a unique doping response has been recently confirmed experimentally.[157-159] Importantly, the accounting for symmetry breaking, the formation of point defects, and using valid computational setups can be used to identify not only false metals but new compounds showing antidoping behavior.

We emphasize again that all of the behaviors seen above: split-off bands, polaron formation, and antidoping are found by mean-field DFT and do not call for special effects such as strong correlation.

## V. When degenerate gapped metals stay metallic

The above examples of false metals demonstrate that real degenerate gapped metals—be that transparent conductors or electrides or Dirac semimetals — might be rare because they can be vulnerable to instabilities converting them to insulators by any of the mechanisms discussed in Sec. III. However, this does not mean that such compounds do not exist and cannot be found. To illustrate it better, let's consider the case of cubic $SrVO_3$ as one of the most widely studied real degenerate gapped metal illustrating why this compound remains metal despite the above-discussed mechanisms. The compound contains light elements, and hence SOC does not affect its electronic structure noticeably. The internal gap in this material is relatively small, which is expected to result in rather small energy lowering due to electron-hole recombination caused by the formation of point defects – indeed, no significant deviation from the ideal composition is observed in $SrVO_3$ under optimized conditions.[160] $SrVO_3$ has closer-to-1 tolerance factor minimizing the need for tilting/rotation and atomic displacements in the compound. When $SrVO_3$ is artificially distorted via tilting/rotation, theory predicts that band gap opening can be archived.[39] However, since such tilting is not energetically favorable, the system is degenerate gapped metal under normal conditions. These results demonstrate how using the classic Goldschmidt tolerance factor can be used as one of the inverse design principles for the search of potential degenerate gapped $ABO_3$ conductors.

The family of real degenerate gapped metals is expanding owing to better understanding resulting from close theory and experiment collaboration. Significant progress has been made in *artificially doped* degenerate gapped compounds, including the classical transparent conducting oxides (e.g., ZnO:Al[161] or $In_2O_3$:Sn[162]) which can accumulate large concentration of free carriers, in the order of $10^{20}$ cm$^{-3}$. Form the theoretical side, the set of heavily doped insulators is expanding with novel prediction of potential n- and p-type transparent conductors[163] with some of the most important systems already experimentally validated (i.e., n-type doped $Ga_2O_3$[164] and $BaSnO_3$[165]). The key steps here remain the understanding of doping limits[25-27] of insulators and identifying main compensation mechanisms. While the *intrinsic* (not doped) degenerate gapped metals have not been the subject of intensive research until recently, the discovery of inorganic electrides[13] and *intrinsic* transparent conductors[24] has raised interest in the field. For instance, a number of new intrinsic degenerate gapped metals have been experimentally synthesized, including 2D $Ca_2N$[166], 1D $Sr_3CrN_3$[167], and 0D $YH_2$[168] electrides. Moreover, many potential gapped metals (e.g., $LaH_2$, $KRb_3$, $Ca_5Bi_3$, $Cs_3O$, $Ba_3CrN_3$, $Ba_3FeN_3$, $Ba_2NaO$, $Li_{12}Mg_3Si_4$, $K_4Al_3(SiO_4)_3$) found in high-throughput calculations of electrides[169,170] and thermoelectrics[171] are awaiting potential laboratory testing: are they stable metals, or will they transform spontaneously to insulators, suggesting that they were false metals, not real metals in the first place?

An interesting property of degenerate gapped metals is the ability of balance the generally conflicting properties of (i) *transparency* (reflecting the internal gap plus the occupied portion of the CB, as well as plasma reflection due to free carriers), (ii) conductivity (reflecting the carriers inside the CB), and (iii) stability (with respect to the metal to insulator transition discussed above). These three factors (i)-(iii) can be used as effective knobs for controlling the properties or real degenerate gapped conductors. For instance, it has been demonstrated that while $BaNbO_3$ and $Ca_6Al_7O_{16}$ are real gapped metals and stable with respect to decompositions to competing phases (i.e., there is a set of experimental conditions corresponding to specific chemical potentials where $BaNbO_3$ and $Ca_6Al_7O_{16}$ can exist), tuning of experimental conditions can result in stabilization of ordered vacancy compounds which have totally different optoelectronic properties[23]. This eventually can open the possibility for tailoring optoelectronic properties by controllable nonstoichiometry to design new functional degenerate gapped metals.

## VI. OUTLOOK AND PERSPECTIVE

The work analyzed and reviewed here points to three relevant possible "call to action" by the community:

(i) *Let's not jump into conclusions that the False Metal Syndrome—incorrect theoretical predictions of degenerate gapped metals based on naïve application of DFT, using the least number of possible magnetic, orbital/structural degrees of freedom – must imply that strong electron-electron correlation is at play, and has been omitted unjustifiably in these calculations*. Indeed, before leapfrogging from N-DFT to highly correlated approaches, its best to examine whether symmetry broken DFT, being a bona fide mean-field theory, contains sufficient physics to explain the absence of false metals. In many cases, the realization that some prediction of metallic states has prematurely motivated the introduction of dynamic electron-electron correlation in a symmetry-preserving picture, as the essential, must-have ingredient.[37,172-174] Such DMFT papers on many $ABO_3$ compounds have explicitly claimed failure of DFT in explaining disproportionation, Jahn–Teller displacements, orbital order, mass enhancement, and gapping in PM phase for many 3d compounds[126,172,174-177], whereas what actually failed was a naïve version of DFT. We pointed out here that there are avenues for removing the constraints on N-DFT theory other than disposing of DFT altogether, leading to a systematic understanding of which compounds are metallic and which are insulating band structures. Such detailed exploration involves considering larger unit cells that support different local structural and magnetic short-range orders. This points to the possible scenario that mean-field like energy lowering via spin- and space- symmetry breaking is the crucial, minimal mechanism at work for systematically explaining metal vs insulating band structures, whereas the consideration the extra physics of dynamic correlation is not needed to describe the actual effects of disproportionation, Jahn–Teller displacements, orbital order, mass enhancement and gapping in 3d oxides phase. Conversely, the "*False Metals syndrome*" needs to be better scrutinized by the DFT community, correcting the naïve approximations in N-DFT that led to it and propagated into databases (Fig. 2).

*(ii) The community should beware of the scenario of predictions of exotic properties in compounds that, in fact, are false metals, i.e., do not exist.*[31,178] The results of N-DFT calculations have been often used as a platform to suggest new exotic physics such as topology, quantum confinement, and superconductivity. Herein, we call for caution regarding proposing compounds with new physical effects on such an uncertain platform (i.e., the compound does not exist). For instance, it has been recently demonstrated that certain topological properties predicted to "live" in compounds that are not realizable due to different energy structural/magnetic symmetry lowering modes.[31,178] As an illustration, recent screening of topological compounds suggested that NiO, $CuBi_2O_4$, $YTiO_3$, $Sr_2IrO_4$, $CaIrO_3$, and $LaTiO_3$ are topological semimetals[20-22], while as demonstrated in Sec. III all the above compounds are false metals becoming real insulators when non-N-DFT is used.

*(iii) Computational band structure databases and high-throughput calculations that recommend compounds as metals but use N-DFT must be scrutinized* (viz. few examples in Fig. 2). This review also revealed the role of computational materials databases[20,43-45] and high-throughput calculations[179-184] to identify metallic vs insulating band structures. Open-access materials databases revolutionized materials science solid-state physics/chemistry field by providing details on new applications of already well-known compounds and a number of potential compounds, which have not been synthesized before. However, the electronic properties of many degenerate gapped metals are still not described correctly within the existing studies (e.g., see examples in Fig. 2). The problem is not easy to handle at present because the databases are in constant change (i.e., hundreds/thousands of new compounds and their properties are added to each database every year), and compounds declared metals one day disappear and are replaced on another day by insulating entries (and this is not a phase transition, just correction of an error) without clear or accessible explanation, leading to confusion due to often weak documentation on the updates. Although it is only a matter of time until the accurate computational setups are commonly used, currently available data for electronic structures for degenerate gapped conductors should be used with great caution and be verified with band gap mechanisms discussed above. The common-sense steps which should be undertaken are: (i) use XC functionals that are able to distinguish occupied from unoccupied states and have reduced SIE; (ii) account SOC especially for high-Z elements; (iii) allow different structural/magnetic symmetry-breaking modalities that can result in energy lowering (e.g., magnetic order, spin-disorder, Jahn–Teller distortion, disproportionation); (iv) verify the stability of compounds with respect to other competing phases and defect formation. All these steps should account for structural nudging and use the cell size not as the fixed input but as the convergence parameter. In these ways, we can demonstrate not only the set of true degenerate gapped metals but also identify the true mechanisms resulting in a change of electronic structures.

**Acknowledgments**

The work on electronic structure calculations was supported by the National Science Foundation, Division of Materials Research, Condensed Matter and Materials Theory (CMMT) within DMR-1724791. The study on doping of quantum materials was supported by U.S. Department of Energy, Office of Science, Basic Energy Sciences, Materials Sciences and Engineering Division within DE-SC0010467. The authors acknowledge the use of computational resources located at the National Renewable Energy Laboratory and sponsored by the Department of Energy's Office of Energy Efficiency and Renewable Energy. This work also utilized the Extreme Science and Engineering Discovery Environment (XSEDE) supercomputer resources, which are supported by the National Science Foundation grant number ACI-1548562. The authors thank Dr. Zhi Wang for useful discussions.

**Data availability**

The data that support the findings of this study are available from the corresponding author upon reasonable request.